\documentclass{aa}  
\usepackage[dvipsnames]{xcolor}
\usepackage[colorlinks=true,allcolors=blue]{hyperref}
\usepackage{graphicx}
\usepackage{txfonts}
\newcommand{\kms}{km\,s$^{-1}$}

\newcommand{\vsini}{$v_{\rm e} \sin i$}

\newcommand{\bz}{$\langle B_{\rm z} \rangle$}
\newcommand{\bs}{$\langle B \rangle$}
\newcommand{\peg}{$o$~Peg}

\newcommand{\figps}[3]{\resizebox{#1}{!}{\rotatebox{#2}{\includegraphics{#3}}}}

\begin{document}

\title{%
A conclusive non-detection of magnetic field in the Am star $o$~Peg with high-precision near-infrared spectroscopy%
%\thanks{Based on observations collected at the ESO Paranal observatory (programme 0111.D-0297).}
\thanks{Based on guaranteed time observations (GTO) collected at the European Southern Observatory (ESO) under ESO programs 0111.D-0297 by the CRIRES$^+$ consortium.}
}
%\subtitle{}

\titlerunning{No magnetic field in $o$~Peg}

\author{
O. Kochukhov\inst{1}
\and A. M. Amarsi\inst{1}
\and A. Lavail\inst{2}
\and H. L. Ruh\inst{3}
\and A. Hahlin\inst{1}
\and A. Hatzes\inst{4}
%\and U. Heiter\inst{1}
\and E. Nagel\inst{3}
\and N. Piskunov\inst{1}
\and K. Pouilly\inst{5}
\and A. Reiners\inst{3}
\and M. Rengel\inst{6}
\and U. Seemann\inst{7,3}
\and D. Shulyak\inst{8}
}

\institute{
Department of Physics and Astronomy, Uppsala University, Box 516, S-75120 Uppsala, Sweden\\
\email{oleg.kochukhov@physics.uu.se}
\and
Institut de Recherche en Astrophysique et Plan\'etologie, Universit\'e de Toulouse, CNRS, IRAP/UMR 5277, 14 avenue Edouard Belin, F-31400, Toulouse, France
\and
Institut f\"ur Astrophysik und Geophysik, Georg-August-Universit\"at, Friedrich-Hund-Platz 1, 37077 G\"ottingen, Germany
\and
Th\"uringer Landessternwarte Tautenburg, Sternwarte 5, 07778 Tautenburg, Germany
\and
Department of Astronomy, University of Geneva, Chemin Pegasi 51, CH-1290 Versoix, Switzerland
\and
Max-Planck-Institut f\"ur Sonnensystemforschung, Justus-von-Liebig-Weg 3, 37077 G\"ottingen, Germany
\and
European Southern Observatory, Karl-Schwarzschild-Str. 2, 85748 Garching bei M\"unchen, Germany
\and
Instituto de Astrof\'isica de Andaluc\'ia – CSIC, Glorieta de la Astronom\'ia s/n, 18008 Granada, Spain
}

\date{Received 29 April 2024 / Accepted 21 May 2024}

\abstract
%context
{
The A-type metallic-line (Am) stars are typically considered to be non-magnetic or possessing very weak sub-G magnetic fields. This view has been repeatedly challenged in the literature, most commonly for the bright hot Am star \peg. Several studies claimed to detect 1--2~kG field of unknown topology in this object, possibly indicating a new process of magnetic field generation in intermediate-mass stars.
}
%aims
{
In this study, we revisit the evidence of a strong magnetic field in \peg\ using new high-resolution spectropolarimetric observations and advanced spectral fitting techniques.
}
%methods
{
The mean magnetic field strength in \peg\ is estimated from the high-precision CRIRES$^{+}$ measurement of near-infrared sulphur lines. This observation is modelled with a polarised radiative transfer code, including treatment of the departures from local thermodynamic equilibrium. In addition, the least-squares deconvolution multi-line technique is employed to derive longitudinal field measurements from archival optical spectropolarimetric observations of this star.
}
%results
{
Our analysis of the near-infrared \ion{S}{i} lines reveals no evidence of Zeeman broadening, ruling out magnetic field with a strength exceeding 260~G. This null result is compatible with the relative intensification of \ion{Fe}{ii} lines in the optical spectrum taking into account blending and uncertain atomic parameters of the relevant diagnostic transitions. Longitudinal field measurements at three different nights also yield null results with a precision of 2~G.
}
%conclusions
{
This study refutes the claims of kG-strength dipolar or tangled magnetic field in \peg. This star is effectively non-magnetic, with the surface magnetic field characteristics no different from those of other Am stars.
}


\maketitle

\section{Introduction}

The upper main sequence stars, with spectral classes from early B to early F, can be separated into two distinct groups according to their surface magnetic field characteristics \citep{preston:1974,donati:2009}. One group, the so-called magnetic chemically peculiar (mCP) or Ap/Bp stars, exhibits stable, globally-organised kG-strength fields on stellar surfaces. These objects comprise about 10\% of all intermediate-mass and massive stars \citep{grunhut:2017,scholler:2017,sikora:2019}. Another, more sizeable, group includes normal and non-magnetic chemically peculiar stars, which show no evidence of strong organised surface magnetic fields. The cool end of the temperature sequence of non-magnetic CP stars is populated by the metallic-line A-type (Am) stars. These stars are known for their slow rotation, common membership in close binary systems, a moderate overabundance of iron-peak elements, and an underabundance of Ca and Sc \citep[e.g.][]{ghazaryan:2018}. The brightest star in the sky, Sirius, is an example of a hot Am star \citep{landstreet:2011,cowley:2016}. 

The spectroscopic characteristics of Am stars facilitate precise measurements of the mean longitudinal magnetic field, \bz, using high-resolution circular polarisation observations \citep{wade:2000}. Applications of this technique, only sensitive to a global magnetic field component, ruled out the presence of $\ga$\,10~G large-scale fields in Am stars \citep{shorlin:2002,auriere:2010a}. This upper limit is at least one order of magnitude below the $\sim$\,100~G weak-field limit of mCP stars \citep{auriere:2007,kochukhov:2023a}. The observational picture of Am-star magnetism becomes more nuanced for weaker fields. Definitive \bz\ measurements at a level of 5--10~G, compatible with a global dipolar-like field sheared by a differential rotation, were reported for at least one Am star, $\gamma$~Gem \citep{blazere:2020}. At the same time, circular polarisation signatures corresponding to sub-G magnetic fields were detected in several bright Am stars \citep{petit:2011,blazere:2016,neiner:2017a}. Considering a conspicuous gap in the magnetic field strength distribution between Am and mCP stars, and different geometrical characteristics of their fields, the ultra-weak Am-star magnetism likely has a different physical origin compared to stable fossil fields found in mCP stars \citep{cantiello:2019,jermyn:2020}.

The view that Am stars are either non-magnetic or host only very weak fields is occasionally challenged in the literature. The star \peg\ (43~Peg, HR~8641, HD~214994) is in the centre of this debate. This is a bright, narrow-line hot Am star, frequently targeted by detailed chemical abundance and model atmosphere analyses based on high-resolution optical \citep{adelman:1988,landstreet:2009,takeda:2012,adelman:2015} and ultra-violet \citep{adelman:1993a} spectra. No longitudinal magnetic field was detected in this star by \citet{shorlin:2002}. Their single magnetic field measurement, \bz\,=\,$-32\pm20$~G, is compatible with zero, albeit at the precision inferior relative to what has been achieved for similarly bright stars in more recent studies. On the other hand, \citet{mathys:1990a} reported the presence of $\approx$\,2~kG mean magnetic field, \bs, in \peg\ from a statistical line-width analysis and an empirical relation between \bs\ and relative equivalent widths of the \ion{Fe}{ii} 614.7 and 614.9~nm lines. \citet{takeda:1991} studied the behaviour of this line pair in a magnetic field using radiative transfer calculations, confirming $\sim$\,2--3~kG magnetic field in \peg. Subsequently, \citet{takeda:1993} measured $\sim$\,2~kG field by minimising the scatter of abundances derived from lines of different strength. These results were recently revised by \citet{takeda:2023}, who analysed new high-quality spectroscopic observations of \peg\ with several of the aforementioned techniques and concluded, once again, that \peg\ possesses \bs\,$\approx$\,1--2~kG.

The vast discrepancy between $\sim$\,kG field strength inferred from the Stokes $I$ spectra and $\sim$\,50~G upper limit obtained from Stokes $V$ observations is usually interpreted assuming that the putative field of \peg\ is ``complex'' or ``tangled'', meaning that it is dominated by a small-scale component that produces broadening and intensification of line profiles in the Stokes $I$ spectra but remains undetectable in Stokes $V$ due to cancellation of the contributions of the surface regions with different field polarities to the disk-integrated stellar polarisation spectra. This situation is often found in cool active stars with dynamo fields \citep[e.g.][]{kochukhov:2020,kochukhov:2023,kochukhov:2021}, although the ratio \bs/\bz\,$\sim$\,100 implied by the previous magnetic field studies \peg\ appears to be exceptionally large even compared to cool stars. Contrary to this picture, \citet{takeda:2023} postulated that \peg\ has a global dipolar magnetic field viewed from the magnetic equator, implying that the null \bz\ measurement by \citet{shorlin:2002} is explained by a fortuitous cancellation of the signals from the positive and negative magnetic hemispheres.

This series of seemingly consistent magnetic field determinations for \peg\ is having a noticeable influence on the community's understanding of Am stars in general. The work by \citet{mathys:1990a} is frequently brought up in review papers \citep{landstreet:1992,dworetsky:1993,smith:1996,kurtz:2000,mathys:2004,mathys:2004a,mathys:2009,hubrig:2021}. It inspired applications of the same magnetic field measurement procedures to other Am stars \citep{lanz:1993a,savanov:1994,scholz:1997a}, to their hotter counterparts, HgMn stars \citep{takada-hidai:1992,hubrig:2001,kochukhov:2013a}, and to other A-type stars \citep{takada-hidai:1993}. In a wider context, some authors consider the presumed complex and strong magnetic field of \peg\ to be representative of all Am stars \citep[e.g.][]{drake:1994}. However, the majority of studies adopt a more conservative viewpoint that this star is, for some reason, exceptional in terms of its magnetic characteristics \citep{debernardi:2000,hui-bon-hoa:2000a,carrier:2002,korcakova:2022}.

In the present study, we revisit the question of whether or not \peg\ is a magnetic star. To this end, we use new spectropolarimetric measurements to characterise the global magnetic field component at a far greater precision than in previous studies. Moreover, we investigate the Zeeman effect using high signal-to-noise (S/N) optical and, for the first time for any Am star, near-infrared high-resolution spectra. The paper is structured as follows. We start with an overview of the observational data employed in this study (Sect.~\ref{sec:obs}). We proceed to describe our polarised spectrum synthesis methodology in Sect.~\ref{sec:method}, including assessment of the deviations from the local thermodynamic equilibrium and partial Paschen-Back (PPB) splitting. This is followed by the analysis of near-infrared (Sect.~\ref{sec:limit}) and optical (Sect.~\ref{sec:pair}) profiles of magnetically sensitive lines, leading to a new sensitive upper limit on the total magnetic field strength of \peg. We complement this analysis with high-precision spectropolarimetric measurements of the longitudinal field (Sect.~\ref{sec:bz}) and derivation of an upper limit of the dipolar magnetic component compatible with these observations. The paper is concluded with the summary and discussion in Sect.~\ref{sec:discus}.

\section{Observations}
\label{sec:obs}

\begin{table*}
\caption{Log of CRIRES$^+$ observations of \peg.}
\label{table:crires-obs}
\centering
\begin{tabular}{c c c c c}     % 5 columns
\hline\hline
UTC date & UTC time at start & Wavelength setting    &   Integration time    & Median S/N per pixel   \\
\hline
2023-08-12 & 05:25:09.087 & Y1029                 & 8x30 seconds          & 372   \\
2023-08-12 & 05:33:12.845 & J1228                 & 8x30 seconds          & 379   \\
2023-08-12 & 05:42:15.479 & H1567                 & 8x30 seconds          & 335   \\
2023-08-12 & 05:51:06.603 & H1582                 & 8x30 seconds          & 309   \\
\hline
\end{tabular}
\end{table*}

\begin{table}[!th]
\centering
\caption{Mean longitudinal magnetic field measurements derived from the ESPaDOnS observations. \label{tbl:bz}}
\begin{tabular}{ccccr}
\hline\hline
UT date & HJD & S/N & S/N$_{\rm LSD}$ & $\langle B_{\rm z} \rangle$ (G) \\
\hline
10-06-2014 & 2456819.122 & 946 & 41391 & $-2.3\pm2.1$ \\
15-06-2014 & 2456824.094 & 938 & 40650 & $-3.1\pm2.2$ \\
19-06-2014 & 2456828.069 & 972 & 41964 & $ 1.5\pm2.1$ \\
\hline
\end{tabular}
\tablefoot{Columns give the UT date and heliocentric Julian date of mid-exposure, the S/N per pixel in the extracted spectrum at 520~nm, the average S/N of the LSD Stokes $V$ profile, and the resulting measurement of the mean longitudinal magnetic field.}
\end{table}

\subsection{Near-infrared spectroscopy}

We observed \peg\ on August 12 2023 with the upgraded CRyogenic
InfraRed Echelle Spectrograph \citep[CRIRES$^+$,][]{dorn:2023}
at the European Southern Observatory (ESO) Very Large Telescope (VLT) located
on Cerro Paranal, Chile. CRIRES$^+$ is a high-resolution
($R \approx 10^5$) cross-dispersed near-infrared spectropolarimeter
mounted on a Nasmyth focus of the 8-m Unit Telescope 3 at the VLT.
Our observations were carried out as part of the CRIRES$^+$ consortium
Guaranteed Time Observations.

The instrument was setup with a slit width of 0.2 arcsecond, and we
obtained observations in four standard wavelength settings: Y1029, 
J1228, H1567, and H1582. 
For observations carried in each wavelength
setting, we obtained 8 individual exposures with a 30-s integration
time, taken with an AAAABBBB nodding pattern. The nodding procedure consists
in placing the star on two distinct positions (A and B) on the slit and
facilitates the removal of sky background and detector artefacts in the
data reduction. The log of our CRIRES$^+$ observations of \peg\ is presented in Table~\ref{table:crires-obs}.

To reduce the data, we used the CRIRES$^+$ data reduction pipeline
\texttt{cr2res} \citep{dorn:2023}. First, we reduced the raw calibration frames
association with our programme taken as part of the daily
calibration routine. These data consist of darks, flat fields, and
wavelength calibration frames (Fabry-Perot etalon, Uranium-Neon lamp)
and were reduced using the standard calibration reduction cascade as
laid out in the \texttt{cr2res} pipeline user manual\footnote{\url{https://www.eso.org/sci/software/pipelines/cr2res/cr2res-pipe-recipes.html
}}.

We then reduced the science data for each wavelength setting using the
\texttt{cr2res\_obs\_nodding} recipe. It applies the reduced
calibrations to the science raw frames (bad pixel mask, pixel-to-pixel
sensitivity), substracts the B frames from the A frames, extracts the 1D
science and error spectra for the two nodding positions using an optimal
extraction algorithm, and applies the wavelength solution to the
spectra.                 

The analysis presented below (Sect.~\ref{sec:limit}) focuses on the \ion{S}{i} lines at 1046~nm observed in the Y1029 setting. The S/N reached at this wavelength is around 480 according to the formal error propagation by the \texttt{cr2res} pipeline. This agrees with the empirical measurement of the standard deviation of continuum points in the vicinity of the \ion{S}{i} lines.

\subsection{Optical spectroscopy}
\label{sec:opt}

Considering particular importance and common historic as well as current usage of the \ion{Fe}{ii} 614.7--614.9~nm line pair for measuring magnetic fields in chemically peculiar stars, including \peg, we revisit this magnetic diagnostic in Sect.~\ref{sec:pair}. At the same time, we refrain from re-assessing all types of magnetic detection methods applied to \peg\ and similar stars based on Stokes $I$ profiles of optical lines since the near-infrared magnetic diagnostic technique developed in our paper (Sect.~\ref{sec:limit}) is by far superior in terms of its ability to detect magnetic fields.

Among multiple optical archival spectra of \peg\ covering this line pair, we chose to use the observed spectrum published by \citet{takeda:2023} due to its higher resolution and better S/N. This spectrum was constructed from a series of individual observations obtained over several nights in October 2008 using the HIDES coud\'e echelle spectrograph at the 1.88~m telescope of Okayama Astrophysical Observatory. The spectrum has a resolving power of $\lambda/\Delta\lambda\approx10^5$ and a very high S/N  approaching 1000. Further details on the acquisition and reduction of these data can be found in \citet{takeda:2012} and \citet{takeda:2023}.

\subsection{Optical spectropolarimetry}

Three previously unpublished circular polarisation observations of \peg\ are included in the PolarBase archive \citep{petit:2014}. These spectra were obtained on three non-consecutive nights between June 10 and June 19, 2014 with the ESPaDOnS spectropolarimeter \citep{donati:2006c} installed at the 3.6~m Canada-France-Hawaii telescope. For each observation, the intensity (Stokes $I$) and circular polarisation (Stokes $V$) spectra were recorded using a sequence of four 110-s sub-exposures, resulting in a peak S/N of 940--970 per pixel of the extracted intensity spectrum at $\lambda=520$~nm. The individual times of mid-exposure and S/N values are provided in Table~\ref{tbl:bz}.

The ESPaDOnS spectra have a fixed format that covers the 370--1030~nm wavelength range with 40 partially overlapping echelle orders at a resolving power of $\lambda/\Delta\lambda=65000$. The \peg\ observations were reduced by the Libre-ESpRIT pipeline \citep{donati:1997} running at the telescope. We post-processed the spectra, aiming to improve continuum normalisation, with the help of the procedures described in \citet{rosen:2018}. In this study, the ESPaDOnS circular polarisation observations are employed to obtain new measurements of the mean longitudinal magnetic field and derive constraints on the global magnetic field geometry in Sect.~\ref{sec:bz}.

\section{Modelling methodology}
\label{sec:method}

\subsection{Polarised spectrum synthesis}

The spectrum synthesis modelling carried out in this paper is based upon the polarised radiative transfer code {\tt Synmast} \citep{kochukhov:2007d,kochukhov:2010a}. This software derives a numerical solution of the polarised radiative transfer equation for a given model atmosphere, limb angle, and local magnetic field vector. {\tt Synmast} shares the equation of state module with the Spectroscopy Made Easy ({\tt SME}) package \citep{piskunov:2017} and has been extensively tested against independent polarised radiative transfer codes \citep{wade:2001}.

In the calculations for the present paper, we adopted a simplified model of stellar magnetic field comprised of a uniform radial field distribution. This approach, widely used in modelling Zeeman broadening and intensification effects in the Stokes $I$ spectra of active stars \citep[e.g.][]{lavail:2019,kochukhov:2020,hahlin:2023}, requires calculation of a small number (typically seven) of local intensity spectra at different limb angles to obtain accurate disk-integrated profiles. Since the radial field is seen from different aspect angles depending on the centre-to-limb position, its cumulative effect on the Zeeman split line profiles is more representative of possible complex magnetic configuration than any field geometry defined relative to the observer's line of sight (e.g. a purely transverse or purely longitudinal field). In fact, the disk-integrated Stokes $I$ line profiles corresponding to a uniform radial magnetic field turn out to be very similar to those calculated for a star covered by an anisotropic randomly oriented field with the same strength. Therefore, our model is compatible with the hypothesis of a complex tangled magnetic field entertained by some previous studies of \peg.

\begin{table*}[!h]
\centering
\caption{Atomic line data adopted for the \ion{S}{i} NIR triplet. \label{tbl:s}}
\begin{tabular}{lccccccccccc}
\hline\hline
Ion & $\lambda$ (nm)$^1$ & $\log gf^2$ & $E_{\rm lo}$ (eV) & \multicolumn{2}{c}{Lower level} & \multicolumn{2}{c}{Upper level} & $g_{\rm eff}$ & $\log \gamma_{\rm rad}$&$\log \gamma_{\rm stark}$& H broadening \\
& & & & $J_{\rm lo}$ & $g_{\rm lo}$ & $J_{\rm up}$ & $g_{\rm up}$ & & & & ($\sigma_{\rm H}/a^2_0$, $\alpha_{\rm H}$)$^3$ \\ 
\hline
\ion{S}{i} & 1045.5451 & $\phantom{-}0.250$ & 6.8601 &1.0&2.0 & 2.0&1.5&1.25&8.95&$-5.37$&625, 0.227\\
\ion{S}{i} & 1045.6757 & $-0.447$ & 6.8601&1.0&2.0 & 0.0&0.0&2.00&8.95&$-5.37$&625, 0.227\\
\ion{S}{i} & 1045.9406 & $\phantom{-}0.030$ & 6.8601 &1.0&2.0 & 1.0&1.5&1.75&8.95&$-5.37$&625, 0.227\\
\hline
\end{tabular}
\tablefoot{(1) NIST, \citet{kramida:2022}; (2) \citet{zerne:1997}; (3) neutral hydrogen broadening cross-sections $\sigma_{\rm H}$ and exponents $\alpha_{\rm H}$ from \citet{barklem:2000b}; other parameters from VALD, \citet{ryabchikova:2015}.}
\end{table*}

\begin{figure*}
\centering
\figps{0.45\hsize}{0}{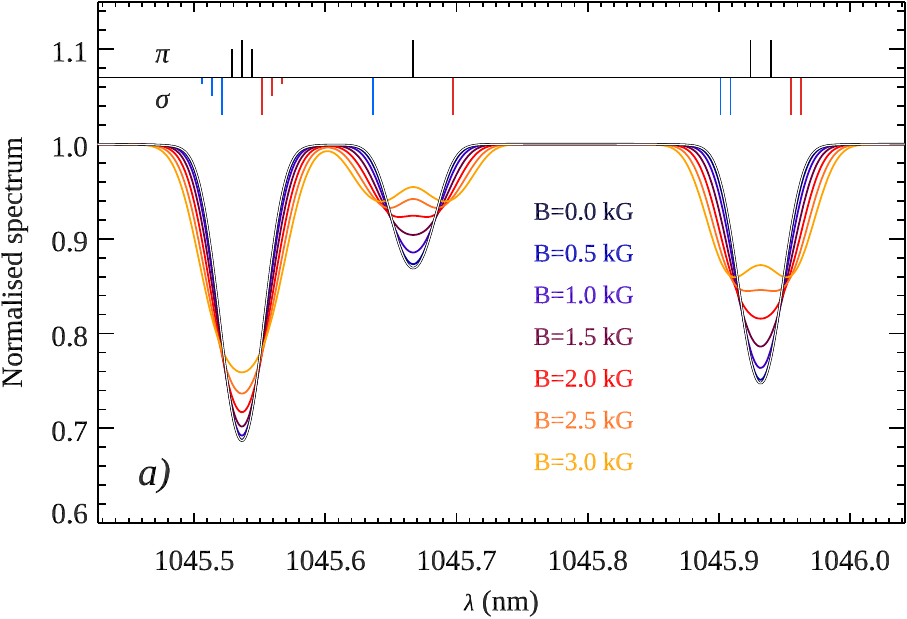} \hspace*{5mm}
\figps{0.45\hsize}{0}{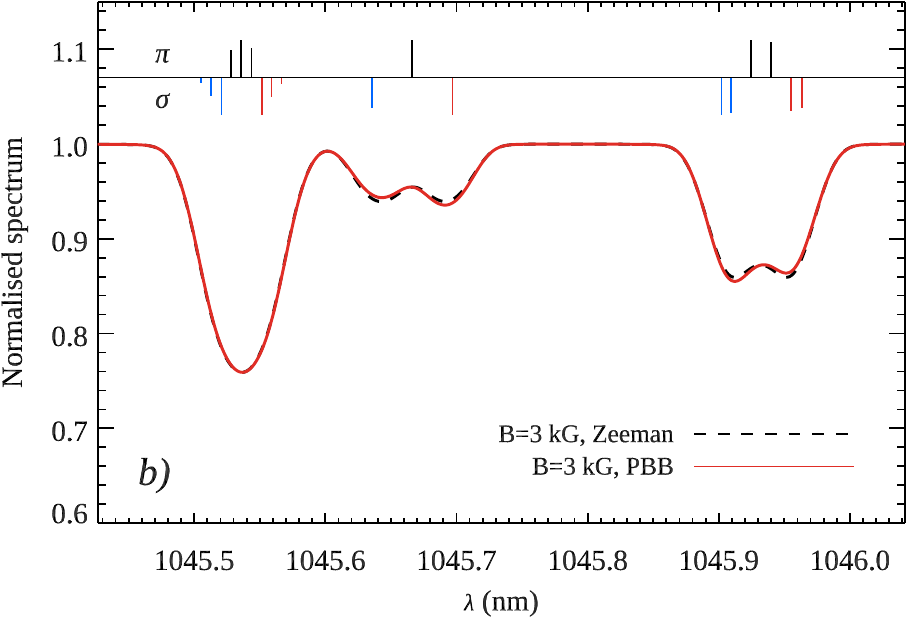}
\caption{{\bf a)} Synthetic Zeeman profiles of the NIR \ion{S}{i} triplet for different magnetic field strengths. The  corresponding splitting patterns are shown schematically above each line for $B=3$~kG. {\bf b)} Comparison of the Zeeman (black dashed line) and PPB (red solid line) calculations for $B=3$~kG. The splitting patterns above line profiles are illustrated for the PPB case.}
\label{fig:s1_zeem}
\end{figure*}

The model atmosphere employed in the present study was calculated with the {\tt LLmodels} code \citep{shulyak:2004} assuming the effective temperature $T_{\rm eff}=9500$~K, surface gravity $\log g=3.6$, and microturbulent velocity $\xi_{\rm t}=2$~\kms. These are typical atmospheric parameter values inferred by modern detailed spectroscopic studies of \peg\ (see the summary in \citealt{takeda:2023}). For these model atmosphere calculations, we used individual chemical abundances of \peg\ from \citet{adelman:2015}.

It is not obvious which spectral lines are best suited for accurate magnetic field measurements in Stokes $I$ in the available wide wavelength coverage optical and NIR spectra of \peg. In particular, spectral lines with the largest effective Land\'e factors are not necessarily the most informative diagnostics of weaker magnetic fields because subtle Zeeman broadening effects are often washed out by rotational broadening. As shown by several studies of cool active stars \citep[e.g.][]{kochukhov:2017c,kochukhov:2020,kochukhov:2023}, spectral lines with above average effective Land\'e factors and complex Zeeman splitting patterns may provide useful alternative diagnostics owing to their strong magnetic intensification response. To identify the most suitable magnetically sensitive lines, we modelled the entire optical and NIR spectrum of \peg\ with {\tt Synmast}, comparing non-magnetic calculations with the theoretical spectra corresponding to $B=1$ and 2~kG. After applying the appropriate rotational and instrumental broadening, the \ion{S}{i} triplet at 1045.5--1045.9~nm has emerged as the most promising diagnostic feature in the NIR wavelength region. The three components of this multiplet are well-separated for the projected equatorial velocity ($v_{\rm e}\sin i$) of \peg\ and exhibit distinct magnetic responses due to different effective Land\'e factors (ranging from 1.25 to 2.00) and diverse Zeeman splitting patterns. The relevant line parameters of the \ion{S}{i} triplet are summarised in Table~\ref{tbl:s}, where we note that the absolute oscillator strengths of the three lines are measured with uncertainties of the order 0.01~dex \citep{zerne:1997}. There are no significant blending contributions to any of these three lines for the effective temperature and chemical abundance of \peg.

Figure~\ref{fig:s1_zeem}a illustrates a series of {\tt Synmast} calculations of the \ion{S}{i} triplet with increasing magnetic field strength. These spectra are convolved with the projected rotational velocity $v_{\rm e}\sin i=5.9$~\kms, macroturbulent velocity $\zeta_{\rm t}=3.7$~\kms\ (similar to the results obtained in Sect.~\ref{sec:limit}), and a Gaussian instrumental profile corresponding to $R=10^5$ to represent instrumental broadening of our CRIRES$^{+}$ spectrum. The effect of magnetic field becomes readily apparent already at $B\approx1$~kG, with the \ion{S}{i} 1045.5~nm line exhibiting less excess magnetic broadening compared to the 1045.9 and, in particular, the 1045.7~nm line. The latter feature starts showing partially resolved Zeeman splitting at $B\approx2$~kG. 
This diverse magnetic response of lines of the same multiplet (or more generally  of nearby lines of the same ion with similar excitation potentials and precise relative oscillator strengths) is ideal for disentangling magnetic broadening and intensification from competing broadening effects due to turbulent and rotational velocity fields. No other spectral lines with the diagnostic potential comparable to the \ion{S}{i} triplet were found in our calculations for the CRIRES$^+$ wavelength region. Hence, we restricted the NIR part of our study to these \ion{S}{i} lines. At the same time, we re-examined the \ion{Fe}{ii} 614.7--614.9~nm lines based on previously published observations (see Sect.~\ref{sec:opt}) to address claims of magnetic field detections using this line pair.

\subsection{Non-LTE effects}

To fully utilise the diagnostic power of the 
\ion{S}{i} 1046~nm triplet, departures  
from local thermodynamic equilibrium (LTE) must be taken into account.
Previous studies of non-LTE effects for sulphur include theoretical investigations by
\citet{2005PASJ...57..751T} and \citet{2009ARep...53..651K},
and applications mainly to the Sun \citep[e.g.][]{2015A&A...573A..25S}
and other FGK-type stars 
\citep[e.g.][]{2007A&A...469..319N,2016A&A...585A..16C},
with fewer studies of warm stars \citep[e.g.][]{2001A&A...375..899K}.
All these studies point to significant effects for the neutral atom.

For this work we carried out our own non-LTE calculations for sulphur.  The
statistical equilibrium was solved for using {\tt Balder}
\citep{2018A&A...615A.139A,2022A&A...668A..68A}. 
The departure coefficients generated by {\tt Balder} were
read into {\tt Synmast} and used to correct the LTE line opacity and source
function prior to spectrum synthesis, as described in \citet{piskunov:2017}.
{\tt Balder} is based on {\tt
Multi3D} \citep{2009ASPC..415...87L} but with several modifications, the most
important for the present work relating to the equation of state and background
opacities (calculated with {\tt Blue}; \citealt{2023A&A...677A..98Z}).  Rayleigh
scattering in the UV was considered for hydrogen \citep{2004MNRAS.347..802L} and
helium \citep{1974PhRvA..10..829L}, while other background transitions were
treated in pure absorption.

\begin{figure}
\centering
\figps{0.9\hsize}{0}{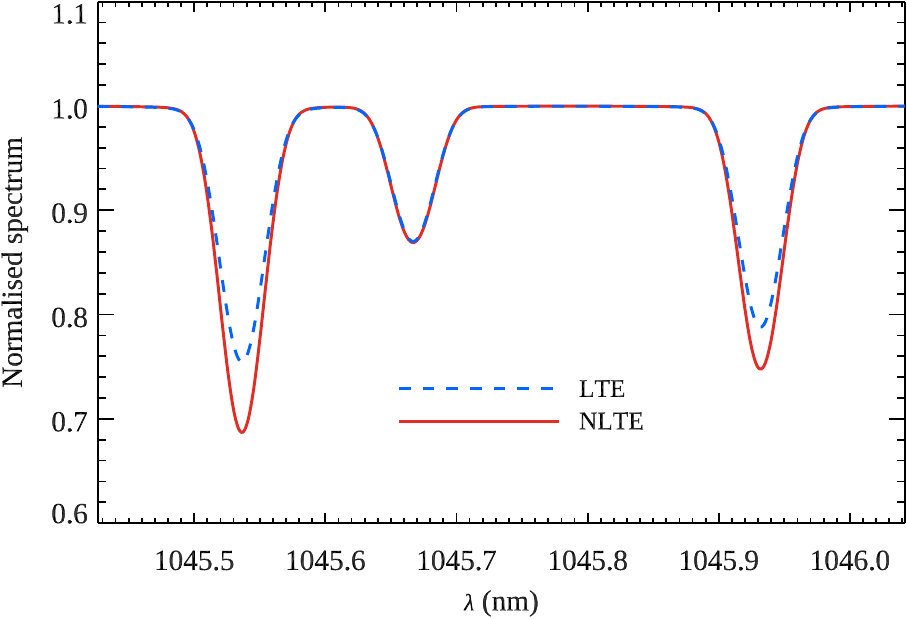}
\caption{Impact of departures from LTE on the synthetic profiles of the NIR \ion{S}{i} triplet. These calculations were carried out for $B=0$~kG, $\log N_{\rm S}/N_{\rm tot}=-4.35$, $\xi_t=2$~\kms\ and other parameters of \peg\ set to the best-fitting values found in Sect.~\ref{sec:limit}.}
\label{fig:s1_nlte}
\end{figure}

The model atom was used in \citet{carlos_submitted} and 
will also be described in a future paper 
(Amarsi et al., in prep.). In brief, the model atom contains
66 levels in total (including seven ``super'' levels of 
neutral sulphur, and four levels of ionised sulphur).
Radiative transition data come from the National Institute of Standards and Technology \citep[NIST,][]{kramida:2022},
based on the calculations of \citet{2006JPhB...39.2861Z};
and from the Opacity Project \citep[e.g.][]{1992RMxAA..23...19S}.
Electron collision data come from empirical
recipes \citep{1962ApJ...136..906V,1973asqu.book.....A}.
Hydrogen collisions were taken from 
\citet{2020ApJ...893...59B} and combined with data calculated
using the recipe of \citet{1991JPhB...24L.127K} in the scattering-length
approximation (see \citealt{2016A&ARv..24....9B} and 
\citealt{2018A&A...616A..89A}).
Pressure broadening by electrons and by neutral hydrogen
was taken into account based on data from
the Kurucz database \citep[e.g.][]{1995ASPC...78..205K}
and by interpolating tables of hydrogen collisional cross-section data 
\citep{1995MNRAS.276..859A,1997MNRAS.290..102B,1998MNRAS.296.1057B}.

The statistical equilibrium calculations were performed on the 
same {\tt LLmodels} atmosphere used for the analysis described
in Sect.~\ref{sec:limit}.
The microturbulent velocity $\xi_{\rm t}$ was fixed
at 2.0~\kms, which is close to the
best-fitting result for \peg\ (see Sect.~\ref{sec:limit}).
Assuming sulphur to be a trace element with no impact on the model
atmosphere, the calculations were performed for 
a range of abundances 
$\log N_{\mathrm{S}}/N_{\mathrm{tot}}=-4.60$ to $-4.25$
($\log N_{\mathrm{S}}/N_{\mathrm{H}}+12=7.42$ to $7.77$),
with step size of $0.05\,\mathrm{dex}$. 

\begin{figure}
\centering
\figps{0.95\hsize}{0}{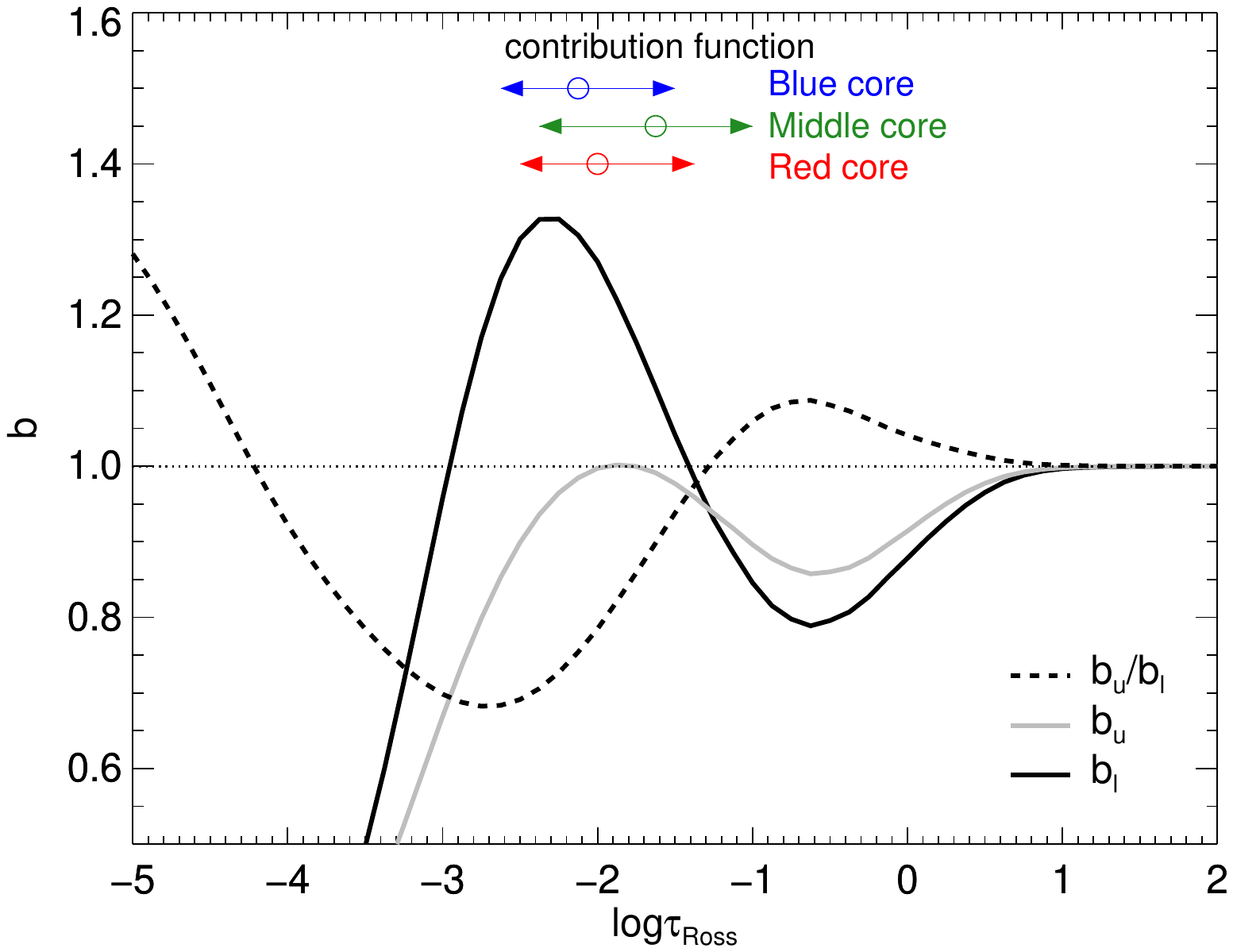}
    \caption{Departure coefficients of the lower 
    and upper levels of the \ion{S}{i} triplet,
    $b_{\text{l}}$ and $b_{\text{u}}$ respectively.
    Their ratio is also shown.
    The full-width at half maximum (arrows) and peak depth (open circles) of the contribution functions to the depression in the disc-integrated flux, calculated at line-centre, are shown for the three lines of the \ion{S}{i} triplet.
}
\label{fig:s1_departures}
\end{figure}

The different components
of the \ion{S}{I} triplet are affected differently by
departures from LTE, as seen in Fig.~\ref{fig:s1_nlte}.
At the best-fitting parameters of \peg,
the strength of the weakest, middle component is quite close
to that in LTE, while the stronger blue and red components are 
significantly stronger in non-LTE. These effects can be understood
from Fig.~\ref{fig:s1_departures}, which shows the departure
coefficients of the lower and upper terms at the best-fitting
abundance as a function of logarithmic Rosseland-mean optical
depth $\log\tau_{\mathrm{Ross}}$.  To first order, the line opacity goes as the departure coefficient 
of the lower level $b_{\text{l}}$, while the line source
function goes as the ratio of the upper and lower level
departure coefficients $b_{\text{u}}/b_{\text{l}}$
\citep[e.g.][]{2003rtsa.book.....R}.
The dip in $b_{\text{l}}$ and $b_{\text{u}}$ in deeper layers,
$\log\tau_{\mathrm{Ross}}\approx-0.6$, is caused by pumping
due to absorption of UV photons in bound-free transitions
(overionisation), resulting primarily in lower line opacity
and a weakening of lines forming in this region.
Towards higher layers, photon losses from high-lying transitions
become significant.  In particular, the \ion{S}{I} triplet itself strongly
regulates the populations of its levels such that
$b_{\text{l}}$ rises and $b_{\text{u}}/b_{\text{l}}$ falls:
they pass through unity at around $\log\tau_{\mathrm{Ross}}\approx-1.4$. 

The different effects on the three components are
directly related to their different formation depths.
This is because the model atom adopts efficient collisional coupling between
the fine structure levels \citep[e.g.][]{2024arXiv240100697L},
and so the departure coefficients are identical for the three components.
The formation depths can be
quantified via the contribution function
to the depression in the disc-integrated flux
\citep[e.g.][]{1996MNRAS.278..337A,2015MNRAS.452.1612A};
these functions calculated at line-centre are
overplotted in Fig.~\ref{fig:s1_departures}.
The stronger blue and red components mostly form 
in layers $\log\tau_{\mathrm{Ross}}\lesssim-1.4$, 
where $b_{\text{l}}>1$ and 
$b_{\text{u}}/b_{\text{l}}<1$: the line opacity is increased
relative to LTE and the line source function is reduced relative to LTE,
and both act to increase the line strength. 
In contrast, the weakest middle component forms in deeper layers,
where it suffers from both line strengthening
and line weakening effects.  Consequently, 
the broadened line profile appears closer to that predicted in LTE.

\begin{figure*}
\sidecaption
\figps{12cm}{0}{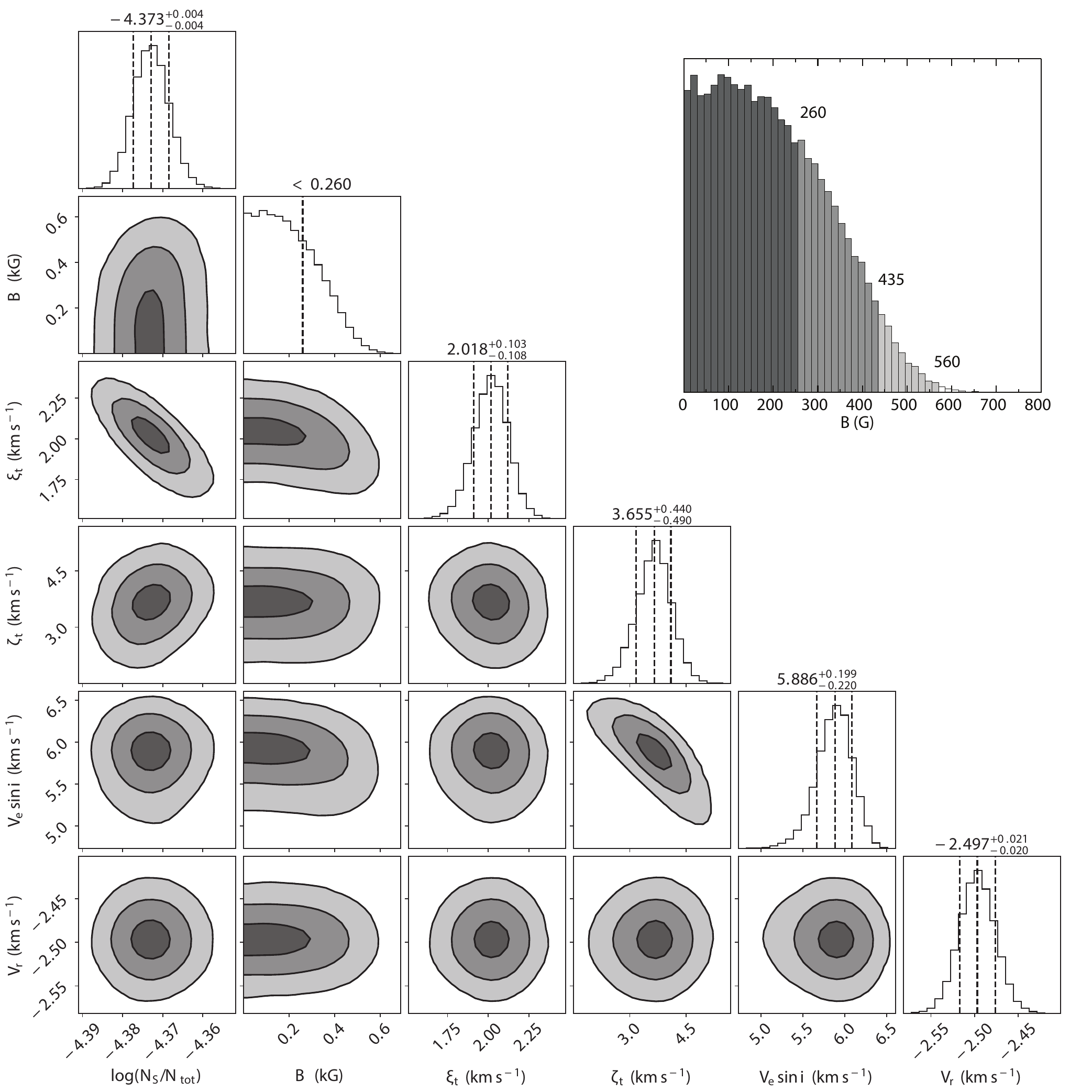}
\caption{Marginalised posterior distributions of the magnetic field strength and other parameters fitted to the NIR \ion{S}{i} triplet. Contours correspond to 1$\sigma$, $2\sigma$, and $3\sigma$ confidence levels. The inset in the upper right corner shows posterior distribution of the field strength with the upper limits corresponding to the same set of confidence levels indicated by different shades of grey.}
\label{fig:corner}
\end{figure*}

\subsection{Partial Paschen-Back splitting}

With the separation of the lower energy levels of $\approx$\,4~cm$^{-1}$, the magnetic splitting of the \ion{Fe}{ii} 614.7--614.9~nm line pair studied in Sect.~\ref{sec:pair} is affected by the PPB effect \citep{mathys:1990}. This phenomenon arises when the magnetic splitting of the atomic energy levels becomes comparable to the fine-structure splitting, resulting in asymmetric line splitting patterns. For lines forming in the PPB regime, both positions and strengths of the Zeeman components exhibit a complex non-linear behaviour with the magnetic field strength. However, deviation from the linear Zeeman splitting in the considered \ion{Fe}{ii} line pair are negligible for fields up to 2--3~kG and their relative equivalent width is essentially unaffected by PPB for $B<4$~kG \citep{takeda:1991}. Thus, in the present study, radiative transfer calculations of the \ion{Fe}{ii} lines did not include the PPB treatment.

The upper energy levels of the \ion{S}{i} triplet studied below differ by 1--2~cm$^{-1}$, also making this group of lines potentially susceptible to PPB. Since no previous computation of the PPB splitting exist in the literature for these lines, we have carried out our own assessment using the methods outlined in \citet{polarization:2004}. This analysis showed that, for the strongest magnetic field considered in Fig.~\ref{fig:s1_zeem}a, the PPB effect produces a detectable distortion of the magnetic splitting patterns, leading to slight asymmetries in the doublet-like splitting of the \ion{S}{i} 1045.7 and 1045.9~nm lines (Fig.~\ref{fig:s1_zeem}b). However, since the subsequent analysis presented in Sect.~\ref{sec:limit} points to a much weaker, if any, magnetic field in \peg, there is no need to incorporate the PPB splitting in our modelling of the \ion{S}{i} triplet.

\section{Analysis of the near-infrared \ion{S}{i} triplet}
\label{sec:limit}

Based on the magnetic spectrum synthesis calculations described above, we modelled the \ion{S}{i} triplet lines in the CRIRES$^+$ spectrum of \peg\ with the help of the Markov chain Monte Carlo (MCMC) technique. For the latter, we employed the {\tt IDL} implementation by \citet{anfinogentov:2021} with a $10^4$ step burn-in followed by $3\cdot10^5$ sampling steps. The free parameters included the non-LTE sulphur abundance $\log N_{\rm S}/N_{\rm tot}$, the microturbulent velocity $\xi_{\rm t}$, the radial-tangential macroturbulent velocity $\zeta_{\rm t}$, the projected rotational velocity $v_{\rm e}\sin i$, the radial velocity shift $v_{\rm r}$, and the magnetic field strength $B$. Uniform priors were assumed for all parameters. The treatment of all three velocity fields as free parameters, without constraints from other lines or previous literature studies, represents a conservative approach likely yielding an increased range of magnetic field strength compatible with observations.

\begin{figure}
\centering
\figps{\hsize}{0}{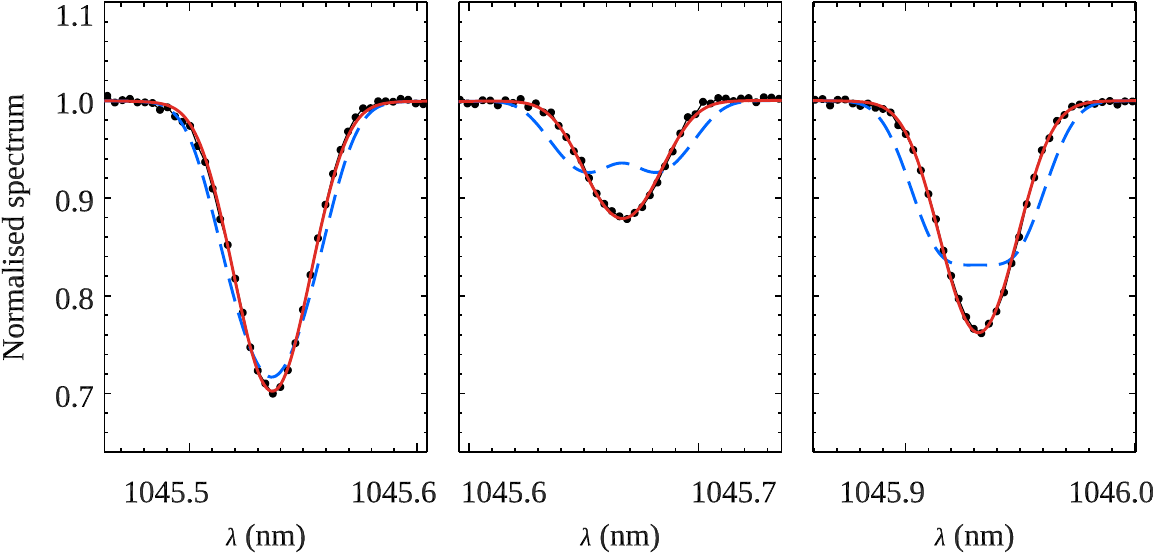}
\caption{Comparison of the \ion{S}{i} lines in the CRIRES$^+$ observation of \peg\ (symbols) with the best-fitting non-magnetic model spectrum (solid lines). The dashed lines show an attempt at fitting the data assuming a 2~kG magnetic field.}
\label{fig:s1_fit}
\end{figure}

The posterior probability distributions resulting from the MCMC evaluation are presented in Fig.~\ref{fig:corner}. It is evident that for most parameters the calculations have converged on Gaussian-like distributions. There is a noticeable correlation between $\zeta_{\rm t}$ and $v_{\rm e}\sin i$, as expected for a narrow-line star. At the same time, the field strength $B$ shows a one-sided distribution extending all the way to $B=0$. Therefore, according to these results, the observed \ion{S}{i} triplet is fully compatible with the null magnetic field. A fit to observations with this model is presented in Fig.~\ref{fig:s1_fit}. The non-magnetic model achieves a near perfect description of the observed profiles. In this case, the standard deviation is 0.36\%, which is close to the observational noise. On the other hand, assuming $B=2$~kG and allowing all other parameters to vary yields theoretical spectra incompatible with observations (dashed line in Fig.~\ref{fig:s1_fit}, standard deviation 2.86\%), even when all other parameters are allowed to vary freely. Repeating the same exercise with theoretical spectra corresponding to $B=1$~kG still yields an unsatisfactory fit (standard deviation 0.51\%).

The final numerical results of the MCMC calculations are summarised in Table~\ref{tbl:mcmc}. For non-zero parameters we report the median value inferred from the corresponding marginalised probability distribution along with the 1-$\sigma$ (13.6\% and 86.4\% percentiles) uncertainties. For the magnetic field strength, the 1-, 2-, and 3-$\sigma$ (68.3, 95.4, and 99.7\% confidence levels, respectively) upper limits are given according to the marginalised probability distribution illustrated in the upper right corner of Fig.~\ref{fig:corner}. The 1-$\sigma$ limit is 260~G, which is by far the smallest magnetic field strength limit estimated from a Stokes $I$ spectrum of an A-type star. The non-magnetic parameters, in particular $\xi_{\rm t}$ and $v_{\rm e}\sin i$, are well within the ranges of determinations by previous studies \citep[e.g.][]{landstreet:2009,gray:2014,adelman:2015}. The non-LTE S abundance derived from the NIR triplet, [S]\,=\,0.54, is also compatible with the measurement of [S]\,=\,$0.39\pm0.10$ from the optical \ion{S}{i} and \ion{}{ii} lines assuming LTE \citep{adelman:2015}.

\begin{table}[!t]
\centering
\caption{Results of MCMC analysis of the \ion{S}{i} NIR triplet. \label{tbl:mcmc}}
\begin{tabular}{lc}
\hline\hline
Parameter & Value \\
\hline
$\log N_{\rm S}/N_{\rm tot}$ & $-4.373\pm0.004$ \\
$\xi_{\rm t}$ (\kms) & $2.02\substack{+0.10\\-0.11}$ \\
$\zeta_{\rm t}$ (\kms) & $3.65\substack{+0.44\\-0.49}$ \\
$v_{\rm e}\sin i$ (\kms) & $5.89\substack{+0.20\\-0.22}$ \\
$v_{\rm r}$ (\kms) & $-2.50\pm0.02$ \\
$B_{1\sigma}$ (G) & $\le260$ \\
$B_{2\sigma}$ (G) & $\le435$ \\
$B_{3\sigma}$ (G) & $\le560$ \\
\hline
\end{tabular}
\tablefoot{The three bottom rows give upper limits of the magnetic field strength corresponding to $1\sigma$, $2\sigma$, and $3\sigma$ confidence intervals.}
\end{table}

\section{Reanalysis of the \ion{Fe}{ii} 614.7--614.9~nm line pair}
\label{sec:pair}

\begin{table*}[!th]
\centering
\caption{Atomic line data adopted for the analysis of the \ion{Fe}{ii} 614.7--614.9 nm line pair. \label{tbl:fe}}
\begin{tabular}{lcccccccccccc}
\hline\hline
Ion & $\lambda$ (nm) & $\log gf$ & $E_{\rm lo}$ (eV) & \multicolumn{2}{c}{Lower level} & \multicolumn{2}{c}{Upper level} & $g_{\rm eff}$ & $\log \gamma_{\rm rad}$&$\log \gamma_{\rm stark}$& \multicolumn{2}{c}{H broadening} \\
& & & & $J_{\rm lo}$ & $g_{\rm lo}$ & $J_{\rm up}$ & $g_{\rm up}$ & & & & $\log \gamma_{\rm vdw}$ & ($\sigma_{\rm H}/a^2_0$, $\alpha_{\rm H}$)$^3$ \\ 
\hline
\ion{Fe}{ii} &614.77341 & $-2.827^1$ & 3.8887 &1.5&1.20 & 0.5 & 2.70 &0.825&8.50&$-6.53$&&186, 0.269\\
\ion{Fe}{i} & 614.78339 & $-1.671^2$  & 4.0758 & 4.0 & 1.26 & 3.0 & 1.21 &1.335&7.55&$-6.03$&$-7.80$& \\
\ion{Fe}{ii} &614.92459 & $-2.841^1$ & 3.8892 &0.5&0.00 & 0.5 & 2.70 &1.350&8.50&$-6.53$&&186, 0.269\\
\hline
\end{tabular}
\tablefoot{(1) \citet{raassen:1998}; (2) \citet{obrian:1991}; (3) \citet{barklem:2000b}; other parameters from VALD, \citet{ryabchikova:2015}.}
\end{table*}

The \ion{Fe}{ii} 614.9~nm line is one of the most commonly used diagnostic lines for detecting Zeeman splitting in the optical spectra of early-type stars. This line has an unusual doublet Zeeman splitting pattern, with two $\pi$ and two $\sigma$ components coinciding in wavelength. This simple splitting pattern enables a straightforward detection of $\ga$\,2~kG magnetic fields in mCP stars and robust measurement of the corresponding mean field modulus \bs\ provided that the stellar projected rotational velocity does not exceed a few \kms\ \citep{mathys:1990,mathys:2017,mathys:1997b}. Additionally, the \ion{Fe}{ii} 614.9~nm line shares the upper energy level with the nearby \ion{Fe}{ii} 614.7~nm transition, which has an almost identical oscillator strength but a very different Zeeman splitting pattern. Similarity of the formation process of these two lines prompted \citet{mathys:1990a,mathys:1992} to suggest using the relative intensification of this line pair, defined using the equivalent widths of these lines as $\delta\equiv2(W_{614.7}-W_{614.9})/(W_{614.7}+W_{614.9})$, as a proxy of the magnetic field strength for stars showing no discernible splitting of the \ion{Fe}{ii} 614.9~nm line. Based on the qualitative comparison of $\delta=0.052$ measured for \peg\ with the relative intensification factors observed for several A stars with different magnetic field strengths, \citet{mathys:1990a} concluded that \peg\ possess a field of $\sim$\,2~kG. This result agrees with $B=2.3$~kG that can be calculated by extrapolating the empirical linear relation between $\delta$ and $B$ calibrated by \citet{mathys:1992} for the 3--5~kG \bs\ interval. The magnetic field diagnostic potential of the \ion{Fe}{ii} 614.7--614.9~nm line pair was further examined by \citet{takeda:1991} using disk-centre polarised radiative transfer calculations and, most recently, by \citet{takeda:2023} assuming an equator-on magnetic field geometry. The former study obtained \bs\,$\approx$\,2--3~kG while the latter one reported \bs\,=\,2.3~kG corresponding to a polar field strength of $B_{\rm d}=3.6$~kG.

These seemingly consistent results notwithstanding, detailed radiative studies of the \ion{Fe}{ii} line pair by \citet{takeda:1991} and \citet{kochukhov:2013a} uncovered several compounding factors and ambiguities in the interpretation of the equivalent width difference of these \ion{Fe}{ii} lines in terms of the magnetic intensification. Both studies demonstrated that, in the field strength range 0--2~kG, the relation between $\delta$ and $B$ is neither linear nor monotonic, rendering the linear relation proposed by \citet{mathys:1992} unusable. Moreover, as emphasised by \citet{kochukhov:2013a}, the $\delta(B)$ relation depends on the assumption about the local magnetic field orientation, with the radial, horizontal, and turbulent magnetic fields yielding different intensification curves. At the same time, these studies did not consider the impact of the choice of atomic parameters of the \ion{Fe}{ii} lines and blending by other spectral features on the derived field strength.

Given the apparent major discrepancy between the tight upper limit of $B$\,$\le$\,260~G from our NIR analysis of \peg\ and repeated claims of 2--3~kG magnetic field from the optical \ion{Fe}{ii} line pair in this star, we re-examined the latter diagnostic with the spectrum synthesis methodology described in Sect.~\ref{sec:method}. The line list adopted for our modelling of the \ion{Fe}{ii} lines is provided in Table~\ref{tbl:fe}. It is based largely on information from VALD, with the oscillator strengths of the \ion{Fe}{ii} lines taken from \citet{raassen:1998}. This choice yields $\Delta\log gf=\log gf_{614.7} - \log gf_{614.9}=0.014$, implying that the \ion{Fe}{ii} 614.7~nm line is slightly stronger than the 614.9~nm line in the absence of magnetic field. Our line list also includes the \ion{Fe}{i} 614.8~nm line, with the experimental $\log gf$ value from \citet{obrian:1991}. Although the importance of this blend for interpretation of the equivalent width of the \ion{Fe}{ii} 614.7~nm line has been mentioned in the literature \citep{mathys:1990a,fossati:2007}, it appears to have been discounted in previous detailed radiative transfer studies of this \ion{Fe}{ii} line pair \citep{takeda:1991,takeda:2023}.

Polarised synthetic spectra were calculated in LTE with {\tt Synmast} for the same atmospheric parameters as in Sect.~\ref{sec:method} assuming a uniform radial magnetic field. Theoretical equivalent widths were obtained with direct integration of line profiles. The same approach was applied to the observed spectra published by \citet{takeda:2023}, confirming his measurement of $\delta=0.025$\footnote{This value is significantly smaller than $\delta=0.052$ reported for \peg\ by \citet{mathys:1990a}. We have no explanation for this discrepancy nor why this reduction of the observed $\delta$ did not result in a corresponding decrease of the field strength inferred by \citet{takeda:2023} compared to \citet{takeda:1991}.}. We estimated the uncertainty of this relative equivalent width measurement to be about 0.004 assuming S/N=1000. 

\begin{figure}
\centering
\figps{0.90\hsize}{0}{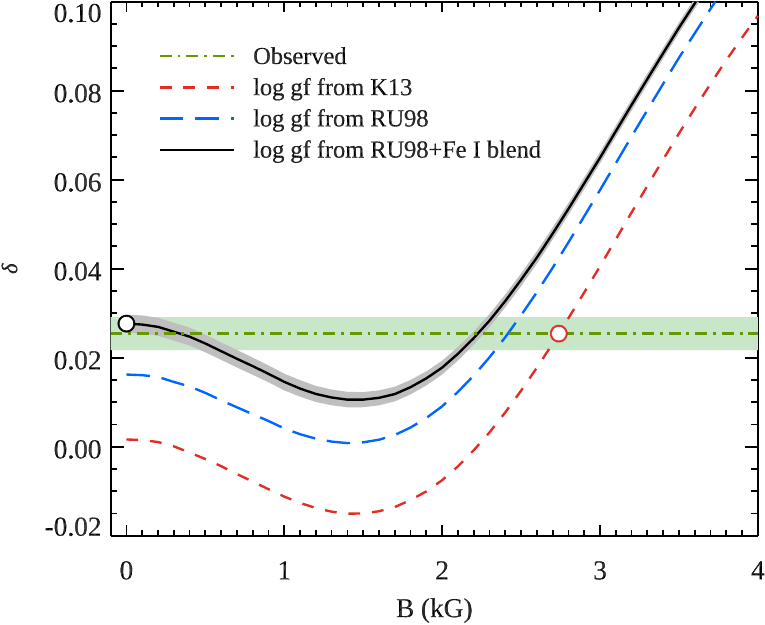}
\caption{Relative intensification of the \ion{Fe}{ii} 614.77 and 614.92 nm line pair as a function of magnetic field strength (solid line with the underlying grey curve illustrating $\pm 1\sigma$ uncertainty arising from the oscillator strength of the \ion{Fe}{i} 614.8~nm blend). The horizontal dash-dotted line corresponds to the observed intensification value with the $\pm 1\sigma$ uncertainty indicated by the green rectangle. The long- and short-dashed lines show theoretical intensification curves for different choices of \ion{Fe}{ii} oscillator strengths (K13: \citealt{kurucz:2013}, RU98: \citealt{raassen:1998}) and blending. The two open circles indicate the line parameter-field strength combinations adopted for the synthetic calculations shown in Fig.~\ref{fig:red_prf}.}
\label{fig:red_ew}
\end{figure}

This observed $\delta$ value is compared with our theoretical $\delta(B)$ curve in Fig.~\ref{fig:red_ew}. Taking the uncertainty of the oscillator strength of the \ion{Fe}{i} blend into account, our calculations reproduce the observed equivalent width difference of the two \ion{Fe}{ii} lines without the need to invoke any magnetic field. The observed $\delta$ is also compatible with our calculations at $B=2.2$~kG, illustrating the ambiguity arising from a non-monotonic $\delta(B)$ relation. We also show in Fig.~\ref{fig:red_ew} another set of calculations in which the \ion{Fe}{i} line was ignored. This shifts the intensification curve downwards, requiring $B=2.4$~kG to reproduce the observed equivalent width difference. Finally, another choice of the \ion{Fe}{ii} oscillator strengths favoured by \citet{takeda:2023}\footnote{We used $\log gf_{614.7}=-2.731$ and $\log gf_{614.9}=-2.732$ according to the semi-empirical calculations by \citet{kurucz:2013}. \citet{takeda:2023} adopted $\log gf_{614.7}=-2.721$ and $\log gf_{614.9}=-2.724$ corresponding to an older version of Kurucz's calculations. In both cases, the resulting $\Delta\log gf$ is smaller than for the oscillator strengths from \citet{raassen:1998}.} yields an even lower $\delta(B)$ curve, requiring a field strength of 2.7~kG to match the observed $\delta$.

\begin{figure}
\centering
\figps{0.95\hsize}{0}{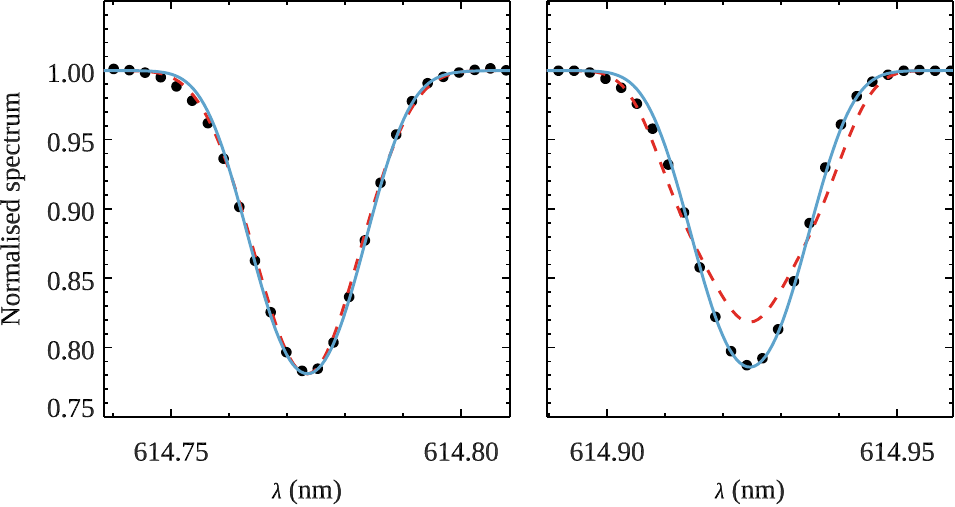}
\caption{Observed (symbols, \citealt{takeda:2023}) and calculated (curves) profiles of the \ion{Fe}{ii} 614.7 and 614.9 nm lines. The solid line shows calculation with the \ion{Fe}{ii} oscillator strengths from \citet{raassen:1998}, blending by \ion{Fe}{i} 614.8 nm line and no magnetic field. The dashed line corresponds to the synthetic spectrum computed with the oscillator strengths from \citet{kurucz:2013}, no \ion{Fe}{i} blending, and $B=2.7$~kG.}
\label{fig:red_prf}
\end{figure}

Thus, a measurement of the magnetic field in \peg\ from equivalent widths of the \ion{Fe}{ii} line pair is generally inconclusive due to ambiguity of the $\delta(B)$ diagnostic and its dependence on the choice of line parameters. Can we extract useful information from the line profiles themselves? Surprisingly, little attention has been paid to this approach despite high-quality observations and theoretical line profile models being readily available. To this end, Fig.~\ref{fig:red_prf} compares the observed profiles published by \citet{takeda:2023} with our {\tt Synmast} calculations underlying the intensification results presented in Fig.~\ref{fig:red_ew}. The non-magnetic theoretical spectrum was calculated with the line list from Table~\ref{tbl:fe}, $\xi_t=2$~\kms, and other stellar parameters ($\log N_{\rm Fe}/N_{\rm tot}$, $\zeta_{\rm t}$, \vsini) adjusted to fit the \ion{Fe}{ii} 614.7~nm line. It is evident that this non-magnetic calculation successfully reproduces the \ion{Fe}{ii} 614.9~nm line without the need to change any line or stellar parameters. Conversely, if we follow \citet{takeda:2023} in adopting the oscillator strengths from \citet{kurucz:2013}, ignore the \ion{Fe}{i} line and use $B=2.7$~kG (required to reproduce the observed $\delta$ for this choice of parameters, see Fig.~\ref{fig:red_ew}), no consistent fit for the two lines can be obtained. In this case, the \ion{Fe}{ii} 614.9~nm line exhibits a reduced central depth and excess broadening compared to \ion{Fe}{ii} 614.7~nm, both incompatible with observations.

To summarise, our re-analysis of the \ion{Fe}{ii} 614.7--614.9~nm line pair does not confirm the presence of 2--3~kG magnetic field in \peg. Instead, the observed equivalent widths and profiles of both lines can be successfully reproduced by non-magnetic spectrum synthesis calculations with a plausible choice of transition probabilites and accounting for the contribution of the \ion{Fe}{i} 614.8~nm blend. The traditional equivalent width-based analysis of this line pair is ambiguous, as both zero and 2--3~kG field strengths yield the same relative equivalent width difference as derived from observations. However, the strong-field solution is ruled out based on the line profile analysis.

\section{Constraints on the global magnetic field}
\label{sec:bz}

There is no evidence of circular polarisation signatures in individual spectral line of \peg\ despite the high quality of ESPaDOnS observations. In this case, similar to other polarimetric studies of weak magnetic fields in intermediate-mass stars \citep[e.g.][]{shorlin:2002,auriere:2010a,makaganiuk:2011a}, we employed the widely used and thoroughly tested least-squares deconvolution \citep[LSD,][]{donati:1997,kochukhov:2010a} procedure to derive very high S/N mean Stokes $I$ and $V$ profiles. This calculation effectively stacks profiles of all suitable metal lines under the assumption that their profiles are self-similar in velocity space and that the Stokes $V$ signal scales with wavelength and effective Land\'e factor of each line according to the weak-field approximation. The line list for LSD profile calculation was obtained from the VALD database \citep{ryabchikova:2015}, using $T_{\rm eff}=9500$~K, $\log g=3.6$ model atmosphere, and elemental abundances from \citet{adelman:2015}. Lines deeper than 5\% of the continuum were retained and wavelength regions affected by the hydrogen Balmer lines or telluric absorption were excluded. The resulting LSD line mask contained 1483 metal lines, dominated by neutral and singly ionised Fe. 

\begin{figure}
\centering
\figps{\hsize}{0}{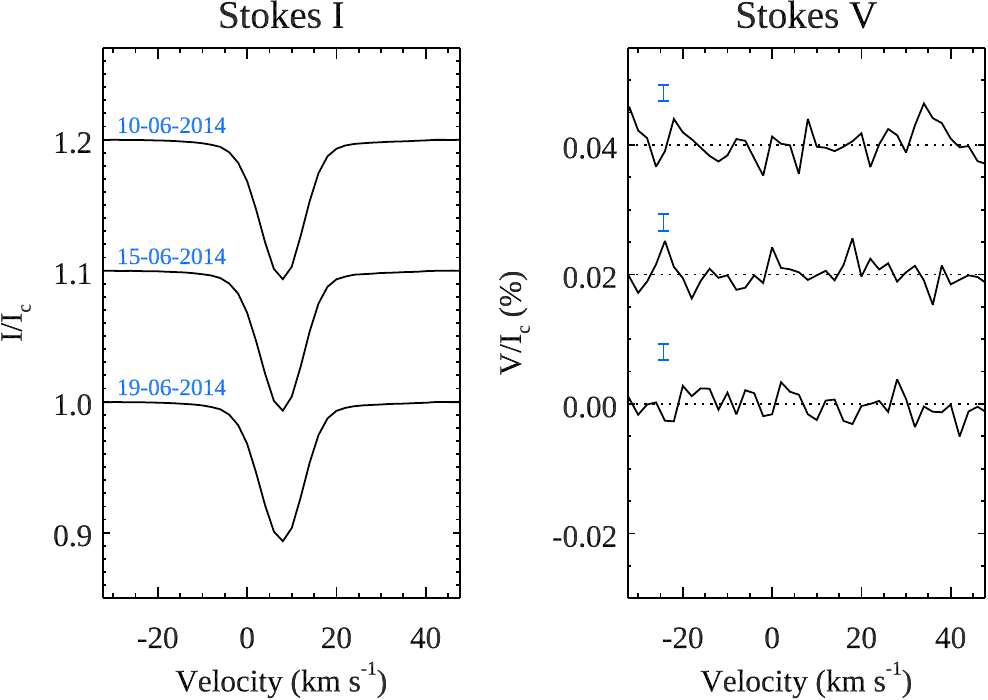}
\caption{LSD profiles of \peg\ derived from archival ESPaDOnS observations. The two panels show Stokes $I$ (left) and $V$ (right) profiles offset vertically for display purposes. The UT observing dates are indicated above each Stokes $I$ profile.}
\label{fig:lsd}
\end{figure}

The LSD intensity and polarisation profiles (the latter normalised to the mean wavelength $\lambda_0=500$~nm and effective Land\'e factor $z_0=1.2$) derived from ESPaDOnS observations are illustrated in Fig.~\ref{fig:lsd}. Application of the LSD procedure resulted in a S/N gain of $\sim$\,40 relative to individual lines, providing a polarimetric precision of $2.4\cdot10^{-5}$. Despite this, no evidence of Stokes $V$ signatures is seen in the Stokes $V$ LSD profiles.

The disk-averaged line-of-sight magnetic field component can be characterised by calculating the mean longitudinal magnetic field \bz\ from the Stokes $I$ and $V$ spectra \citep[e.g.][]{mathys:1991,bagnulo:2002a}. Here we used the prescription by \citet{wade:2000} and \citet{kochukhov:2010a} for deriving \bz\ from LSD profiles. The resulting measurements are reported in the last column of Table~\ref{tbl:bz}. We achieve no detection of \bz, with all three measurements consistent with zero field at the precision of 2.1--2.2~G.

Analysis of LSD profiles derived from high-resolution circular polarisation observations allows us to constrain global magnetic field topologies which produce null \bz, for example toroidal field or equator-on dipole. At the same time, one very specific global field configuration -- a dipole closely aligned with the stellar rotational axis observed from the stellar equator -- will yield no signal in Stokes $V$ since spectral contributions from the two stellar hemispheres will cancel out exactly for all rotational phases. This type of global magnetic field geometry, with a polar field of $B_{\rm d}=2$--4~kG, was advocated by \citet{takeda:2023} to reconcile the absence of polarimetric magnetic detections in \peg\ with the existence of strong field according to his analysis of intensity spectra. However, our new high-precision polarimetric measurements suggest that this magnetic configuration is highly improbable. Using the LSD profile modelling procedure described in \citet{metcalfe:2019}, we established that it is sufficient for the magnetic obliquity angle $\beta$ to deviate by 0.16\degr\ from 0\degr\ or 180\degr\ or for the stellar inclination angle $i$ to deviate by the same amount from 90\degr\ for a 3~kG dipolar field to produce Stokes $V$ signatures incompatible with observations at 3-$\sigma$ confidence level. Assuming random orientations of the rotational and magnetic axes, the probability that both $i$ and $\beta$ angles fall in the intervals corresponding to no detectable polarimetric signatures is $<$\,$10^{-8}$. Thus, the modern high-quality polarimetric data analysed here essentially rule out the possibility that \peg\ harbours a strong large-scale magnetic field. 

\section{Summary and discussion}
\label{sec:discus}

In this study, we revisited the question of the presence of a strong magnetic field at the surfaces of metallic-line stars. We focused on the bright Am star \peg, for which the existence of 1--2 kG field covering the entire stellar surface was suggested by studies spanning over three decades \citep{mathys:1990a,lanz:1993a,takeda:1991,takeda:1993,takeda:2023}. Owing to this series of investigations, \peg\ is often considered as a prototypical Am star with a strong complex magnetic field. We re-examined archival optical spectra available for \peg\ and acquired new high-resolution and high S/N near-infrared spectroscopic observations using CRIRES$^+$ at the ESO VLT. With the help of these NIR observations, we identified the \ion{S}{i} triplet at 1046~nm as a particularly useful diagnostic for detecting Zeeman broadening signatures of magnetic field in \peg\ given its large sulphur overabundance. Modelling these lines with a non-LTE polarised radiative transfer code coupled with a Bayesian inference framework, we concluded that magnetic field of \peg\ cannot exceed 260~G (1-$\sigma$ confidence level). This represents the most sensitive magnetic field upper limit derived from the intensity spectra of an Am star.

Further re-analysis of the \ion{Fe}{ii} 614.7--614.9~nm line pair, which was previously used to claim magnetic field in \peg, revealed importance of the choice of atomic line parameters and accounting for weak blending. We showed that the observed relative intensification of these lines can be reproduced with either non-magnetic calculations or $>$\,2~kG magnetic field, depending on the adopted line list. However, the profile shape, in particular that of the 614.9~nm \ion{Fe}{ii} line, is incompatible with the strong-field solution. To summarise, both the optical and NIR spectroscopic data suggests that \peg\ does not possess a kG-strength magnetic field. If any surface field is present in this star, its strength does not exceed a few hundred G.

The outcome of our re-assessment of the magnetic interpretation of the relative intensification of the \ion{Fe}{ii} 614.7--614.9~nm lines in \peg\ is reminiscent of the results of investigations of these lines in the spectra of HgMn stars. For these objects, often considered hotter analogues of Am stars, complex kG-strength magnetic fields claimed from the \ion{Fe}{ii} line strengths \citep{hubrig:1999a,hubrig:2001} were shown to be incompatible with the lack of Zeeman broadening in high-resolution spectra \citep{kochukhov:2013a}. The line intensification analysis was also similarly jeopardised by blending in some HgMn targets \citep{takada-hidai:1992}. This experience for two distinct classes of CP stars suggests that using the relative intensification of this \ion{Fe}{ii} line pair to infer magnetic field strength tends to yield spurious results when extrapolated outside the empirically calibrated 3--5~kG field strength interval \citep{mathys:1992}. Therefore, some of the recent magnetic field strength estimates employing the \ion{Fe}{ii} line ratio \citep[e.g.][]{holdsworth:2014,holdsworth:2016,smalley:2015,murphy:2020} may not be trustworthy, even when applied to the types of CP stars expected to host strong magnetic fields.

In addition to the analysis of Zeeman broadening and magnetic intensification in Stokes $I$ spectra, we obtained complementary information on the global magnetic field of \peg\ using unpublished archival optical circular polarisation spectra collected with ESPaDOnS at CFHT. Three observations spread over 10 nights yield null mean longitudinal magnetic field measurements with a typical uncertainty of 2~G. This represents a 10-fold precision improvement compared to the previous \bz\ determinations for \peg\ \citep{shorlin:2002,bychkov:2009}. Moreover, we reported no evidence of polarisation signatures in S/N\,$\approx$\,40000 LSD Stokes $V$ profiles, ruling out the presence of moderately complex non-dipolar global magnetic field topologies.

The non-detection of a magnetic field in polarimetric observables does not necessarily rule out that the star hosts a sizeable surface magnetic field. It can be argued, as was done in previous studies of \peg, that the star possesses a dipolar field that happened to be observed nearly equator-on \citep{takeda:2023} or that the stellar surface field lacks any global component \citep{mathys:1990a}. However, a high polarimetric accuracy of the Stokes $V$ LSD profiles derived in our study challenges both of these hypotheses. To remain consistent with the polarimetric non-detections in all three Stokes $V$ observations, the stellar and dipolar field axes must have a very particular orientation. Namely, both the inclination and magnetic obliquity angles must not deviate by more than $\approx$\,0.2\degr\ from 90\degr\ and 0\degr\ or 180\degr, respectively. We showed that the probability to encounter such a magnetic configuration is negligible. On the other hand, the second hypothesis of a highly tangled magnetic field implies the ratio of the total to global field strength of $\sim$\,$10^3$, if one adopts the $\sim$\,2~kG total field strength advocated by previous studies of \peg. Although the situations of \bz\,$\ll$\,\bs\ are not unheard of in stellar magnetometry, this ratio appears to be extremely large for \peg\ compared to $10^1$--$10^2$ total to global field amplitudes found in cool active stars with dynamo fields \citep{kochukhov:2020,kochukhov:2021}. Thus, to accommodate the low \bz\ and high \bs\ for \peg, one would require a qualitatively new type of magnetic field generation process producing a highly structured surface field with a degree of complexity not seen in any other types of stars. However, our results point to a more prosaic explanation: the previous claims of 1--2~kG complex field in \peg\ are erroneous and the actual total field strength does not exceed a few hundred G according to the most sensitive NIR diagnostic. Therefore, this star should be currently considered non-magnetic according to the most sensitive polarimetric and spectroscopic magnetic field detection analyses.

Taking into account results of the present study as well as recent literature, there is no evidence that either globally organised or complex kG-strength magnetic fields exist on the surfaces of Am stars. Instead, Am stars observed with sufficient polarimetric precision appear to possess ultra-weak sub-G fields \citep{petit:2011,blazere:2016,neiner:2017a} of unknown geometry. Thus, our work reaffirms the presence of a ``magnetic desert'' -- a conspicuous two orders of magnitude gap in the field strength distribution between the 100--300~G lower field strength bound of mCP stars \citep{auriere:2007,kochukhov:2023a} and the sub-G fields of Am stars. The only CP star known to straddle this gap is the marginal Am star $\gamma$~Gem, which hosts a dipolar field with a polar strength of $\approx$\,30~G \citep{blazere:2020}. It remains to be seen if this object is part of a larger population or represents an extreme example of mCP star, which is in the process of loosing its fossil magnetic field.

\begin{acknowledgements}
O.K. acknowledges support by the Swedish Research Council (grant agreements no. 2019-03548 and 2023-03667), the Swedish National Space Agency, and the Royal Swedish Academy of Sciences.
A.M.A. acknowledges support from the Swedish Research Council (grant agreement no. 2020-03940).
H.L.R. acknowledges the support of the DFG priority program SPP 1992 ``Exploring the Diversity of Extrasolar Planets (RE 1664/20-1)''.
E.N. acknowledges the support by the DFG Research Unit FOR2544 ``Blue Planets around Red Stars''.
K.P. acknowledges the Swiss National Science Foundation, grant number 217195, for financial support.
M.R. acknowledges the support by the DFG priority program SPP 1992 ``Exploring the Diversity of Extrasolar Planets'' (DFG PR 36 24602/41).
D.S. acknowledges financial support from the project PID2021-126365NB-C21 funded by Agencia Estatal de Investigación of the Ministerio de Ciencia e Innovación (MCIN/AEI/10.12039/501100011033).
CRIRES$^+$ is an ESO upgrade project carried out by Th\"uringer Landessternwarte Tautenburg, Georg-August Universit\"at
G\"ottingen, and Uppsala University. The project is funded by the Federal Ministry of Education and Research (Germany) through
Grants 05A11MG3, 05A14MG4, 05A17MG2 and the Wallenberg Foundation.
\end{acknowledgements}

%\bibliographystyle{aa}
%\bibliography{astro_papers,ama}

\end{document}